\documentclass[aps,twocolumn,showpacs,amsmath]{revtex4}
\usepackage{graphicx}
\usepackage{bm}
\newcommand{\ave}[1]{\langle #1\rangle}
\newcommand{\ket}[1]{{\left|{#1}\right\rangle}}
\newcommand{\bra}[1]{{\left\langle{#1}\right |}}
\newcommand{\oper}[1]{\hat{{#1}}}
\newcommand{\sinc}{{\mathrm{sinc}}}
\newcommand{\qv}{{\bm{\mathrm{q}}}}
\newcommand{\kv}{{\bm{\mathrm{k}}}}
\newcommand{\rv}{{\bm{\mathrm{r}}}}
\begin{document}
\title{Spatial antibunching of photons with parametric down-conversion}
\author{ W.A. T. Nogueira}
\author{S. P. Walborn}
\author{S. P\'adua}
\author{C. H. Monken}\email{monken@fisica.ufmg.br}
\affiliation{Departamento de F\'{\i}sica,
Universidade Federal de Minas Gerais, Caixa Postal 702,
Belo Horizonte, MG 30123-970, Brazil}

\date{\today}

\begin{abstract}
The theoretical framework behind a recent experiment by Nogueira
{\it{et al.}} [Phys.  Rev.  Lett.  {\bf{86}}, 4009 (2001)] of spatial
antibunching in a two-photon state generated by collinear type II
parametric down-conversion and a birefringent double-slit is
presented.  The fourth-order quantum correlation function is evaluated
and shown to violate the classical Schwarz-type inequality, ensuring
that the field does not have a classical analog.  We expect these
results to be useful in the rapidly growing fields of quantum imaging
and quantum information.
\end{abstract}


\maketitle

\section{Introduction}
\label{sec:intro}
As current technology advances, more and more attention is placed upon
Quantum Mechanics to solve future problems.  Furthermore, quantum
systems are capable of performing some tasks more efficiently than
classical systems \cite{chuang00}, drawing even more emphasis to
quantum technologies.  In particular, the fields of optical
communication, optical imaging and optical information processing have
been appended by the rapidly developing fields of quantum
communication \cite{barbosa98,enk99,pereira00}, quantum imaging
\cite{gatti99,abouraddy01} and quantum information processing
\cite{chuang00}.  Thus, the study of quantum phenomena promises to be
a fruitful enterprise.

For many years researchers have studied the non-classical behavior of
light, such as squeezing \cite{stoler70,stoler71,slusher85} and
antibunching \cite{carmichael76,kimble76,kimble77}.  However, most
theoretical and experimental investigations deal with time variables
only.  That is, most treatments consider only one spatial mode.  In a
recent review article, Kolobov \cite{kolobov99} demonstrates that many
quantum phenomena also occur when considering spatial variables of the
electromagnetic field.  Many areas of technology stand to benefit from
the possible applications provided by such quantum phenomena.

An invaluable tool in these areas of research is the generation of
entangled photons using parametric down-conversion \cite{burnham70}.
The two-photon state of light exhibits non-separable behavior
\cite{fonseca99a,fonseca01} and has been used in nearly all quantum
information schemes \cite{cabello01}.

Spatial antibunching was recently observed experimentally by Nogueira
{\em{et al.}} \cite{nogueira01} using spontaneous parametric
down-conversion (SPDC).  In this article, we provide a theoretical
background for the experiment reported in \cite{nogueira01}.  Section
\ref{sec:antib} is dedicated to the general introduction of temporal
and spatial antibunching.  In section \ref{sec:spdc} we discuss the
theoretical observation of spatial antibunching of photons using a
two-photon entangled state produced by SPDC, as in \cite{nogueira01}.
We close with some concluding remarks in section \ref{sec:conc}.

\section{Photon bunching and antibunching}
\label{sec:antib}
It is well known that any state of the electromagnetic field that has
a classical analog can be described by means of a positive nonsingular
Glauber-Sudarshan $P$ distribution, which has the properties of a
classical probability functional over an ensemble of coherent states.
Because of this fact, the normally-ordered intensity correlation
function for stationary fields must obey the following inequality
\cite{mandel95}:
\begin{equation}
\ave{{\mathcal T\!}:\hat{I}({\rv},t)\hat{I}({\rv},t+\tau):} \leq
\ave{:\hat{I}^{2}({\rv},t):},
\label{eq:in1q}
\end{equation}
where ${\mathcal T\!}:\  :$ stands for time and normal ordering.
Photon density operators are defined as
\begin{equation}
\hat{I}({\rv},t)=\hat{\bm{V}}^{\dagger}({\rv},t)\cdot\hat{\bm{V}}({\rv},t),
\end{equation}
where
\begin{equation}
\hat{\bm{V}}({\rv},t)=\frac{1}{\sqrt{\Omega}}\sum_{{\kv},\sigma}
\hat{a}_{{\kv},\sigma}\bm{\epsilon}_{{\kv},\sigma}
e^{i({\kv}\cdot{\rv}-\omega_{\kv} t)},
\label{eq:vdef}
\end{equation}
$\hat{a}_{{\kv},\sigma}$ is the annihilation operator for the
mode with wave vector ${\kv}$ and polarization $\sigma$,
$\bm{\epsilon}_{{\kv},\sigma}$ is the unit polarization
vector, $\Omega$ is the quantization volume and $\omega = c k$.

Expression (\ref{eq:in1q}) is commonly written in the
shorter form
\begin{equation}
G^{(2,2)}(\rv_{1},\rv_{2},\tau)\le G^{(2,2)}(\rv_{1},\rv_{2},0),
\label{eq:short}
\end{equation}
where
\begin{equation}
G^{(2,2)}(\rv_{1},\rv_{2},\tau)= \ave{{\mathcal
T\!}:\hat{I}({\rv},t)\hat{I}({\rv},t+\tau):}.
\end{equation}
Since the delayed photon coincidence detection probability
$\mathcal{P}(\rv_{1},\rv_{2},\tau)$ is proportional to
$G^{(2,2)}(\rv_{1},\rv_{2},\tau)$ \cite{mandel95}, inequality
(\ref{eq:short}) means that for the class of fields considered above,
photons are detected either bunched or randomly distributed in time.
Photon antibunching in time, characterized by the violation of
(\ref{eq:in1q}), was predicted by Carmichael and Walls
\cite{carmichael76}, Kimble and Mandel \cite{kimble76}, and was first
observed by Kimble, Dagenais and Mandel in resonance fluorescence
\cite{kimble77}.

In the space domain, the concept analogous to stationarity is
homogeneity.  For a homogeneous field, the expectation value of any
quantity that is a function of position is invariant under translation
of the origin \cite{mandel95}.  In particular, on a plane surface
normal to the propagation direction,
\begin{equation}
G^{(2,2)}(\bm{\rho}_{1},\bm{\rho}_{2},\tau)=
G^{(2,2)}(\bm{\delta},\tau)
\label{eq:stat1}
\end{equation}
and
\begin{equation}
\ave{:I^{n}(\bm{\rho}+\bm{\delta},t+\tau):} =
\ave{:I^{n}(\bm{\rho},t):},
\label{eq:uniform}
\end{equation}
where $\bm{\rho}$ is the transverse position vector,
$\bm{\delta}=\bm{\rho}_{1}-\bm{\rho}_{2}$ and $n=1,2,\ldots$

For homogeneous and stationary fields described by positive
nonsingular $P$ distributions, the Schwarz inequality implies that
\begin{equation}
\ave{\mathcal{T}\!:\hat{I}(\bm{\rho},t)\hat{I}(\bm{\rho}+\bm{\delta},t+\tau):}
\leq \ave{:\hat{I}^{2}(\bm{\rho},t):},
\label{eq:cs2q}
\end{equation}
that is,
\begin{equation}
G^{(2,2)}(\bm{\delta},\tau)\leq
G^{(2,2)}(\bm{0},0).
\label{eq:bunch}
\end{equation}
Analogously to what was concluded from inequality (\ref{eq:short}),
for fields that admit classical stochastic models, inequality
(\ref{eq:bunch}) implies that photons are detected either spatially
bunched or randomly spaced in a transverse detection screen. 
Violation of (\ref{eq:bunch}) implicates the possibility of quantum
fields exhibiting spatial antibunching.  Spatial antibunching of
photons has been predicted by some authors
\cite{kolobov99,berre-rousseau79,klyshko82,bialynicka-birula91,kolobov91}.

\section{Spatial antibunching with down-conversion}
\label{sec:spdc}
\begin{figure}
\centerline{\includegraphics{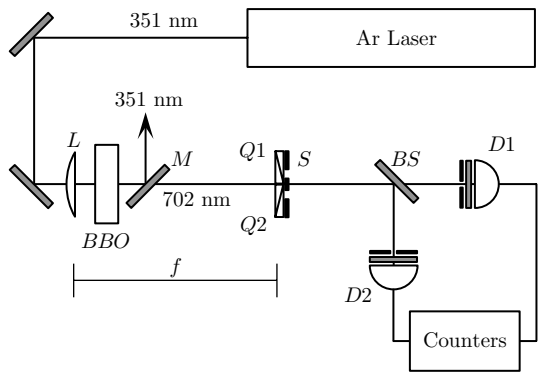}}
  \caption{ Schematic diagram of spatial antibunching setup.  An Ar
  laser pumps a BBO crystal, generating correlated photons.  The
  down-converted photons are incident on the birefringent double slit S
  and then the beamsplitter $BS$.  The pump beam is focused on the
  double slit.  Single and coincidence counts are registered with
  detectors $D_{1}$ and $D_{2}$.}
\label{fig:exp}
\end{figure}

In this section we show that a field that violates inequality
(\ref{eq:bunch}) can be generated by means of spontaneous parametric
down-conversion.  The experimental setup we are considering is shown
in Fig. \ref{fig:exp}.  A nonlinear birefringent crystal is used to
generate collinear entangled photon pairs.  The down-converted photons
are then incident on a birefringent double-slit (see section
\ref{subsec:bds}) and coincidences are detected by detectors $D_{1}$
and $D_{2}$.  The pump beam is focused on the center of the plane of
the double-slit, between the two slits.  Interference
filters are used such that the monochromatic approximation is valid.

\begin{figure}
\centerline{\includegraphics{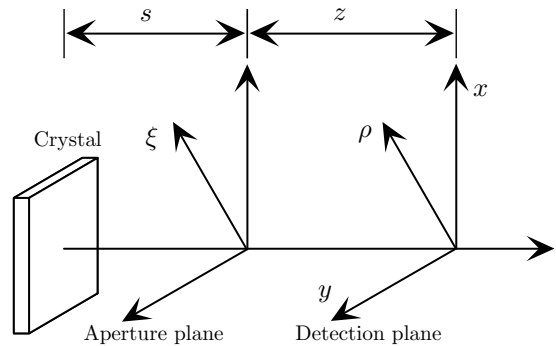}}
\caption{Illustration of the geometry. $s$ is the crystal--aperture distance
and $z$ is the aperture--detector distance.}
\label{fig:geom}
\end{figure}

The following discussion refers to the basic geometry illustrated in
Fig. \ref{fig:geom}, where a thin crystal is separated from an
aperture plane by a distance $s$ and the aperture plane is separated
from a detection plane by a distance $z$.

Using a treatment based in reference \cite{hong85}, in the paraxial
and monochromatic approximations, collinear  SPDC generates a
quantum state of the form \cite{monken98a}:
\begin{equation}
\ket{\psi}_{\mathrm{SPDC}}=C_{1}\ket{\mathrm{vac}}+C_{2}\ket{\psi}
\label{eq:13}
\end{equation}
with
\begin{equation}
\ket{\psi}=\int\hspace{-2mm}\int\limits_{D}\hspace{-1mm} d\qv_{1}
d\qv_{2}\Phi(\qv_{1},\qv_{2})\ket{\qv_{1},\sigma_{1}}
\ket{\qv_{2},\sigma_{2}}.
\label{eq:14}
\end{equation}
The coefficients $C_1$ and $C_2$ are such that $|C_{2}| \ll \,
|C_{1}|$.  $C_2$ depends on the crystal length, the nonlinearity
coefficient, the magnitude of the pump field, among other factors. 
The kets $\ket{\qv_{j},\sigma_{j}}$ represent Fock states labeled by
the transverse component $\qv_{j}$ of the wave vector $\kv_{j}$ and
the polarization $\sigma_{j}$ of the down-converted photon $j=1,2$. 
In this paper we consider type-II phase matching, in which case
$\sigma_{1}=e$ and $\sigma_{2}=o$ where $e$ $(o)$ stands for
extraordinary (ordinary) polarization.  $\ket{\psi}$ is the two-photon
component of the total quantum state.  The function
$\Phi(\qv_{1},\qv_{2})$, which can be regarded as the normalized
angular spectrum of the two-photon field \cite{monken98a}, is given by
\begin{equation}
\Phi(\qv_{1},\qv_{2}) = \sqrt{\frac{2L}{\pi^2K}}\
v(\qv_{1}+\qv_{2})\ \sinc\left(\frac{L|\qv_{1}-\qv_{2}|^{2}}{4K}
\right),
\label{eq:15}
\end{equation}
where $v(\qv)$ is the normalized angular spectrum of the pump beam,
$L$ is the length of the nonlinear crystal in the $z$-direction, and
$K$ is the magnitude of the pump field wave vector.  The integration
domain $D$ is, in principle, defined by the conditions $q_{1}^{2}\le
k_{1}^{2}$ and $q_{2}^{2}\le k_{2}^{2}$.  However, in most
experimental conditions, the domain in which $\Phi(\qv_{1},\qv_{2})$
is appreciable is much smaller than that.  The state written above is
not to be considered as a general expression for the SPDC process. 
Its validity is determined by experimental conditions, especially by
the detection apparatus.  As long as the monochromatic and paraxial
approximations are valid, the results predicted by expression
(\ref{eq:14}) are in excellent agreement with experience. 
Monochromatic approximation is guaranteed by the presence of
narrow-band interference filters in the detection apertures, whereas
paraxial approximation is guaranteed by keeping transverse detection
regions much smaller than their distance from the crystal.


We consider for now that the down-converted fields are incident on
some sort of aperture, so as to produce fourth-order interference in
the absence of second-order interference.  The reason for such a
requirement is the following: Spatial photon antibunching is a
fourth-order effect in a homogeneous field, that is to say, in a field
that, according to \ref{eq:uniform}, does not show intensity patterns. 
With this scheme, we are seeking for a fourth-order interference
pattern that depends only on $x_{1} - x_{2}$, the relative position of
detectors.  Furthermore, this fourth-order interference pattern must
have a minimum when $x_{1} = x_{2}$ in order to produce antibunching. 
Fourth-order spatial interference in the absence of second-order can
be achieved in spontaneous parametric down-conversion by means of a
double-slit whose slit separation is much greater than the transverse
coherence length of the down-converted field, as reported by Fonseca
\textit{et al.} \cite{fonseca99}.  However, in reference
\cite{fonseca99} the fourth-order correlation function, which is
proportional to the coincidence rate, depends on $x_{1} + x_{2}$
instead of $x_{1} - x_{2}$.  In order to achieve a minimum of
coincidences when $x_{1} = x_{2}$, we have to introduce a phase
difference of $\pi$ between the two possibilities\\ \{photon 1 through
slit 1, photon 2 through slit 2\} and \{photon 1 through slit 2,
photon 2 through slit 1\}.  In our experiment, that phase difference
was introduced by means of birefringent elements placed in front of
each slit, as described later.  After the aperture, the two-photon
state can be written as
\begin{eqnarray}
\ket{\psi}&=&M\hspace{-1mm}\sum\limits_{\sigma_{1}^{\prime},
\sigma_{2}^{\prime}} \int\hspace{-2mm}\int\hspace{-2mm}\int\hspace{-2mm}\int
\hspace{-1mm} d\qv_{1} d\qv_{2}d\qv_{1}^{\prime}
d\qv_{2}^{\prime}\ \Phi_{\!A}(\qv_{1},\qv_{2}) \nonumber \\
& &\times
T_{\sigma_{1}\sigma_{1}^{\prime}}(\qv_{1}^{\prime}-\qv_{1})\
T_{\sigma_{2}\sigma_{2}^{\prime}}(\qv_{2}^{\prime}-\qv_{2})\nonumber\\
& &\times\ket{\qv_{1}^{\prime},\sigma_{1}^{\prime}}
\ket{\qv_{2}^{\prime},\sigma_{2}^{\prime}},
\label{eq:scatt}
\end{eqnarray}
where $M$ is a normalization constant,
$\Phi_{\!A}(\qv_{1},\qv_{2})$ is the angular spectrum of the biphoton
field on the aperture plane, that is,
\begin{eqnarray}
\Phi_{\!A}(\qv_{1},\qv_{2})&=&\sqrt{\frac{2L}{\pi^2K}}\ v(\qv_{1}+\qv_{2})\
\sinc\left(\frac{L}{4K}|\qv_{1}-\qv_{2}|^{2} \right)\nonumber\\
& &\times\exp\!\left[i\ s\left(k_{1}+k_{2}-\frac{q_{1}^{2}}{2k_{1}}-
\frac{q_{2}^{2}}{2k_{2}}\right)\right].\nonumber\\ 
\label{eq:phiaa}
\end{eqnarray}
$T_{\sigma\sigma^{\prime}}(\qv)$ is the transfer function of the
aperture, linking
the incident field with transverse wave vector $\qv$ and polarization
$\sigma$ with the scattered field with transverse wave vector
$\qv^{\prime}$ and polarization $\sigma^{\prime}$.
$T_{\sigma\sigma^{\prime}}(\qv)$ is given by the Fourier
transform of the aperture function
$A_{\sigma\sigma^{\prime}}(\bm{\xi})$.

Since we are working with collinear SPDC with
$k_{1}=k_{2}=\frac{1}{2}K$, $\Phi_{\!A}$ is written as
\begin{eqnarray}
\Phi_{\!A}(\qv_{1},\qv_{2})&=&\sqrt{\frac{2L}{\pi^2K}}\ v(\qv_{1}+\qv_{2})\
\sinc\left(\frac{L}{4K}|\qv_{1}-\qv_{2}|^{2} \right)\nonumber\\
& &\times \exp\left[\frac{-i\ s}{2K}\left(|\qv_{1}+\qv_{2}|^{2}
+|\qv_{1}-\qv_{2}|^{2}\right)\right],\nonumber\\
\label{eq:phia}
\end{eqnarray}
where the irrelevant phase factor $e^{iKs}$ is omitted.

Using the orthonormal properties of the Fock states, we can define
\begin{equation}
\bm{\Psi}(\bm{\rho}_{1},\bm{\rho}_{2}) =
\bra{\mathrm{vac}}\oper{\bm{V}}(\bm{\rho}_{2})\otimes
\oper{\bm{V}}(\bm{\rho}_{1})\ket{\psi}
\label{eq:amplidef}
\end{equation}
as the two photon coincidence detection amplitude, where
\begin{equation}
\oper{\bm{V}}(\bm{\rho})=e^{ikz}\sum\limits_{\sigma}\int
d\qv\ \oper{a}_{\sigma}(\qv)\
\bm{\epsilon}_{\sigma}\ e^{i(\qv\cdot\bm{\rho}-
\frac{q^{2}}{2k}z)}
\label{eq:vparax}
\end{equation}
is the monochromatic form of (\ref{eq:vdef}) in the paraxial
approximation and $z$ is the distance between the aperture plane and
the detection plane, as shown in Fig. \ref{fig:geom}.  It is assumed
that the polarization vector $\bm{\epsilon}$ is independent of $\qv$.
The two-photon coincidence-detection probability for stationary fields
is proportional to the fourth-order correlation function with
$\tau=0$:
\begin{equation}
  {\mathcal{P}}(\bm{\rho}_{1},\bm{\rho}_{2}) \propto
G^{(2,2)}(\bm{\rho}_{1},\bm{\rho}_{2},0) =
||\bm{\Psi}(\bm{\rho}_{1},\bm{\rho}_{2})||^{2}.
\label{eq:probdetect}
\end{equation}
\subsection{The birefringent double-slit}
\label{subsec:bds}

\begin{figure}
\centerline{\includegraphics{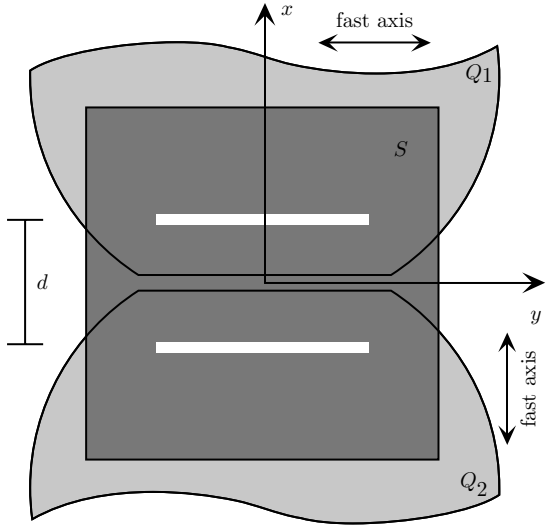}}
\caption{The birefringent double slit.  The quarter wave
plates $Q_{1}$ and $Q_{2}$ are aligned with orthogonal fast axes.  $S$
is a double slit with slits separation $d$.}
\label{fig:slit}
\end{figure}

The birefringent double slit consists of two quarter-wave plates
mounted in front of a typical double slit, such that each wave plate
covers only one slit and their fast axes are orthogonal to one
another, as shown in Fig. \ref{fig:slit}.  The slits are separated a
distance $d$.  With the plate-slit aperture oriented such that the
slits and one fast axis are parallel to the $e$ ($y$) direction and
the other fast axis parallel to the $o$ ($x$) direction, we can
approximate the field-aperture functions by
\begin{eqnarray}
A_{oo}(\bm{\xi}) & = &
-i\delta(\xi_{x}-d/2)+\delta(\xi_{x}+d/2)
\nonumber \\ A_{ee}(\bm{\xi}) & = &
\delta(\xi_{x}-d/2)-i\delta(\xi_{x}+d/2)
\nonumber \\ A_{eo}(\bm{\xi}) & = & 0 \nonumber \\
A_{oe}(\bm{\xi}) & = & 0,
\label{eq:slits}
\end{eqnarray}
where $\xi_{x}$ is the $x$-component of $\bm{\xi}$.  The plate-slit
apertures provide a controlled phase factor, that is, no phase will be
added to a field with polarization parallel to the direction of the
fast-axis of the wave-plate, while a field with perpendicular
polarization will be modified by a phase factor of $\exp(-i\pi/2)$.
Thus, the phase factor depends on the polarization of the field as
well as through which slit the field ``passes". 

\subsection{The coincidence-detection probability amplitude}
Combining equations (\ref{eq:scatt}--\ref{eq:vparax}) and
(\ref{eq:slits}), we arrive at the following expression for
coincidence-detection amplitude in the Fraunhofer approximation:
\begin{equation}
\bm{\Psi} = \Psi_{eo}\ [\bm{\epsilon}_{e}\otimes \bm{\epsilon}_{o}] +
\Psi_{oe}\ [\bm{\epsilon}_{o}\otimes \bm{\epsilon}_{e}],
\end{equation}
where
\begin{eqnarray}
\Psi_{\sigma_{1}\sigma_{2}} &\propto  & \int\hspace{-2mm}\int
\hspace{-1mm} d\qv_{1}d\qv_{2}\
\Phi_{\!A}(\qv_{1},\qv_{2}) \qquad\qquad\nonumber \\
& & \times\left\{ \cos\left[ \frac{d}{2}(q_{1x} + q_{2x})-\frac{kd}{2z}(x_{1} +
x_{2})\right]\right.\nonumber\\
& & \left. \pm\sin\left[\frac{d}{2}(q_{1x} - q_{2x})-\frac{kd}{2z}(x_{1} -
x_{2})\right]\right\},\nonumber\\
\label{eq:detect}
\end{eqnarray}
where the ``$+$'' holds for $\Psi_{eo}$ and the ``$-$'' holds for
$\Psi_{oe}$.

We assume that the pump field is a gaussian beam whose waist is
located on the aperture plane:
\begin{equation}
u_{A}(\bm{\xi}) \propto e^{-{\xi}^{2}/w_{0}^{2}}.
\label{eq:28}
\end{equation}
Its angular spectrum is
\begin{equation}
v_{A}(\qv) = v(\qv)\ \exp\!{\left[i\ 
s\left(K-\frac{q^{2}}{2K}\right)\right]} \propto
e^{-w_{0}^{2}{q}^{2}/4},
\label{eq:laser}
\end{equation}
where $w_{0}$ is the radius of the beam waist.  Using (\ref{eq:phia})
and (\ref{eq:laser}) in (\ref{eq:detect}), it is straightforward to
show that
\begin{equation}
\Psi_{\sigma_{1}\sigma_{2}} \propto
  e^{-(\frac{d}{2w_{0}})^{2}}\cos\!\left[\frac{kd}{2z}(x_{1} +
x_{2})\right]\!\mp\sin\!\left[\frac{kd}{2z}(x_{1} -
x_{2})\right]
\label{eq:detect2}
\end{equation}
It is interesting to note that the length $L$ of the nonlinear crystal
enters in the coincidence-detection amplitude only as a multiplicative
constant.  It is clear from expression (\ref{eq:detect2}) above, that
the fulfillment of the homogeneity condition (\ref{eq:uniform}) for
$n=2$ in the $x$-direction depends on the factor
$e^{-(\frac{d}{2w_{0}})^{2}}$.  If $w_{0}\ll d$, the dependence on
$x_{1}+x_{2}$ disappears and transverse field on the detection plane
can be considered as homogeneous.  This is the reason why the pump
beam must be focused on the center of the double slit.  In this case,
\begin{equation}
{\Psi}_{eo}(\bm{\rho}_{1},\bm{\rho}_{2})= - 
{\Psi}_{oe}(\bm{\rho}_{1},\bm{\rho}_{2})\propto
\sin\left[\frac{kd}{2z}(x_{1}-x_{2})\right].
\label{eq:30}
\end{equation}
Thus, the coincidence detection probability is
\begin{equation}
{\mathcal{P}}(\bm{\rho}_{1},\bm{\rho}_{2}) \propto
1-\cos\left[\frac{kd}{z}(x_{1}-x_{2})\right].
\label{eq:31}
\end{equation} 
When $x_{1}=x_{2}$, the coincidence count rate is zero and increases
with $x_{1} - x_{2}$ until $(x_{1}-x_{2}){kd}/{z}=\pm\pi/2$.
Therefore, the fourth-order correlation function
$G^{(2,2)}(\bm{\rho}_{1},\bm{\rho}_{2},t)$, which is
proportional to the coincidence detection probability
${\mathcal{P}}(\bm{\rho}_{1},\bm{\rho}_{2})$, does not have a maximum
at $x_{1}=x_{2}$.  This contradicts (\ref{eq:bunch}), thus characterizing
spatial antibunching of photons. 

\section{Discussion and conclusion}
\label{sec:conc}
We have shown the theoretical background behind the spatial
antibunching of photons using parametric down-conversion.  It may be
instructive for the reader to compare the experiment analyzed here
with its classical counterpart.  In this context, the single count
detection rate $R_{cl}(x)$ should be proportional to the classical
average intensity $\ave{I(x)}$, whereas the coincidence count rate
$C_{cl}(x_1,x_2)$ should be proportional to the intensity-intensity
(or the fourth-order) correlation function $\ave{I(x_1)I(x_2)}$.  The
single count detection rate of down-converted light in the presence of
a double-slit has been studied in previous works
\cite{fonseca99,ribeiro94,ribeiro99}.  In reference \cite{ribeiro94}
it was demonstrated that in terms of its single count rate, SPDC
behaves like a classical Schell-model extended light source.  In our
experiment, the transverse coherence length being shorter than the
slits separation and shorter than the slits widths themselves, the
single count rate is given by the classical expression for incoherent
illumination, which can be approximated by a gaussian 
\begin{equation}
R_{cl}(x)\propto e^{-\frac{x^2}{2\sigma^2}}, 
\end{equation} 
where $\sigma = \frac{z}{ka}$ and $a$ is the width of the slits. 
Since the transverse detection range $x_{max}-x_{min}$ is much shorter
than the width of this gaussian profile for the slits-detectors
distance considered, the single count rate is fairly constant over de
detection range \cite{nogueira01}.  By another side, the coincidence
detection rate due to a classical source is totally different from the
observed with down-converted light.  Perhaps, the best classical model
for type-II SPDC is a superposition of two extended light sources
orthogonally polarized and correlated in intensity.  After the light
is diffracted by the birefringent double-slit, the calculation of the
classical fourth-order correlation function is quite similar to the
case of the Hanbury Brown -- Twiss intensity interferometer
\cite{hbt}.  Classical intensity interferometry is known to be
insensitive to phase.  Therefore, the birefringent elements have no
effect on the predicted fourth-order correlation, that is,
\begin{equation} 
C_{cl}(x_1,x_2)\propto
1+v\cos\left[\frac{kd}{z}(x_{1}-x_{2})\right].  
\label{eq:hbt}
\end{equation} 
The visibility $v$ is in the range $0\leq v \leq \frac{1}{2}$ and
depends on the statistics of the source.  It is clear from expression
(\ref{eq:hbt}) above that $C_{cl}(x_1,x_2)$ predicts spatial bunching,
as expected from any classical light source.  In view of the above
analysis, the results presented here describe an entirely quantum
fourth-order interference effect, with no classical analog
\cite{belinsky92}.  In addition to rendering further interest in the
study of non-classical states of light, spatial antibunching promises
to be a useful tool in quantum imaging and quantum information
technologies.

\begin{acknowledgments}
The authors acknowledge financial support from the Brazilian agencies
CNPq and CAPES.
\end{acknowledgments}


\begin{thebibliography}{99}

\bibitem{chuang00}
M.~A. Nielsen and I.~L. Chuang, {\em Quantum Computation and Quantum
Information} (Cambridge, Cambridge, 2000).

\bibitem{barbosa98}
G.~A. Barbosa, Phys.  Rev.  A \textbf{58}, 3332 (1998).

\bibitem{enk99}
S.~J. van Enk, H.~J. Kimble, J.~I. Cirac and P. Zoller, Phys.  Rev.  A
\textbf{59}, 2659 (1999).

\bibitem{pereira00}
S.~F. Pereira, Z.~Y. Ou and H.~J. Kimble, Phys.  Rev.  A \textbf{62},
042311 (2000).

\bibitem{gatti99}
A. Gatti, E. Brambilla, L. Lugiato and M.~I. Kolobov, Phys.  Rev.
Lett.  \textbf{83}, 1763 (1999).

\bibitem{abouraddy01}
A.~F. Abouraddy,B.~E.~A. Saleh, A.~V. Sergienko and M.~C. Teich, Phys.
Rev.  Lett.  \textbf{87}, 123602 (2001).

\bibitem{stoler70}
D. Stoler, Phys.  Rev.  D. \textbf{1}, 3217 (1970).

\bibitem{stoler71}
D. Stoler, Phys.  Rev.  D. \textbf{4}, 1925 (1971).

\bibitem{slusher85}
R. Slusher, L. Hollberg, B. Yurke, J. Mertz and J. Valley, Phys.  Rev.
Lett.  \textbf{55}, 2409 (1985).

\bibitem{carmichael76}
H.~J. Carmichael and F.~F. Walls, J. Phys.  B. \textbf{9}, L43 (1976).

\bibitem{kimble76}
H.~J. Kimble and L. Mandel, Phys.  Rev.  A. \textbf{13}, 2123 (1976).

\bibitem{kimble77}
H.~J. Kimble, M. Dagenais and L. Mandel, Phys.  Rev.  Lett.  \textbf{39},
691 (1977).

\bibitem{kolobov99}
M.~I. Kolobov, Rev.  Mod.  Phys.  \textbf{71}, 1539 (1999).

\bibitem{burnham70}
D, Burnham and D. Weinberg, Phys.  Rev.  Lett.  \textbf{25}, 84 (1970).

\bibitem{fonseca99a}
E. J. S. Fonseca, C. H. Monken and S. P\'adua, Phys.  Rev.  Lett.  \textbf{82},
2868 (1999).

\bibitem{fonseca01}
E. Fonseca, Z. Paulini, P. Nussenzveig, C. Monken and S. P\'adua,
Phys.  Rev.  A. \textbf{63}, 043819 (2001).

\bibitem{cabello01}
See reference [1] and the ``Bibliographic guide to the foundations of
quantum mechanics and quantum information'' by A. Cabello,
{\tt{quant-ph/0012089}}.

\bibitem{nogueira01}
W.~A.~T. Nogueira, S.~P. Walborn, S. P\'adua and C.~H. Monken , Phys.
Rev.  Lett.  \textbf{86}, 4009 (2001).

\bibitem{mandel95}
L. Mandel and E. Wolf, {\em Optical Coherence and Quantum Optics}
(Cambridge University Press, New York, 1995).

\bibitem{berre-rousseau79}
M.~L. Berre-Rousseau, E. Ressayre and A. Tallet , Phys.  Rev.  Lett.
\textbf{43}, 1314 (1979).

\bibitem{klyshko82}
D.~N. Klyshko, Sov.  Phys.  JETP. \textbf{56}, 753 (1982).

\bibitem{bialynicka-birula91}
Z. Bialynicka-Birula, I. Bialynicki-Birula and G.~M. Salamone, Phys.
Rev.  A. \textbf{43}, 3696 (1991).

\bibitem{kolobov91}
M.~I. Kolobov and I. Sokolov, Europhys.  Lett.  \textbf{15}, 271 (1991).

\bibitem{hong85}
C.~K. Hong and L. Mandel, Phys.  Rev.  A. \textbf{31}, 2409 (1985).

\bibitem{monken98a}
C.~H. Monken and P.~H.~S. Ribeiro and S. P\'adua, Phys.  Rev.  A. 
{\bf57}, 3123 (1998).

\bibitem{fonseca99}
E. J. S. Fonseca, C. H. Monken, S. P\'adua, and G. A. Barbosa, Phys. 
Rev.  A \textbf{59}, 1608 (1999).  

\bibitem{ribeiro94} 
P. H. S. Ribeiro, C. H. Monken and G. A. Barbosa, Appl.  Opt.  
\textbf{33}, 352 (1994).

\bibitem{ribeiro99} 
P. H. Souto Ribeiro, S. P\'{a}dua and C. H. Monken Phys.  rev.  A
{\bf60}, 5074 (1999).


\bibitem{hbt}
R. Hanbury Brown, {\em The Intensity Interferometer} (Taylor \&
Francis, London, 1974).

\bibitem{belinsky92}
A.~V. Belinsky and D.~N. Klyshko, Phys.  Lett.  A. \textbf{166}, 303
(1992).


\end{thebibliography}


\end{document}